%% file: main.tex
\documentclass{siamart220329}

\usepackage{tikz}
\usetikzlibrary{patterns}
\usetikzlibrary{positioning}
\usetikzlibrary{calc}
\usetikzlibrary{arrows.meta}
\usetikzlibrary{shadings}

\usepackage[
  per-mode=symbol,
  bracket-unit-denominator = false
]{siunitx}

\usepackage{booktabs}

\usepackage{amsmath, amsfonts}
\usepackage[labelformat=simple]{subcaption}

\usepackage{algorithmic}

\newcommand{\head}[1]{\raggedleft{#1}}

\title{Algorithms for Parallel Shared-Memory Sparse Matrix-Vector Multiplication on Unstructured Matrices}
\author{Kobe Bergmans\thanks{Department of Computer Science, KU Leuven, Leuven, 3001, Belgium (\email{kobe.bergmans@kuleuven.be}, \email{karl.meerbergen@kuleuven.be}, \email{raf.vandebril@kuleuven.be}).}
\and Karl Meerbergen\footnotemark[1]
\and Raf Vandebril\footnotemark[1]}

\headers{Algorithms for Sparse Matrix-Vector Multiplication}{K. Bergmans, K. Meerbergen, R. Vandebril}

\begin{document}
\maketitle

\begin{abstract}
    The sparse matrix-vector (SpMV) multiplication is an important computational kernel, but it is notoriously difficult to execute efficiently. This paper investigates algorithm performance for unstructured sparse matrices, which are more common than ever because of the trend towards large-scale data collection. The development of an SpMV multiplication algorithm for this type of data is hard due to two factors. First, parallel load balancing issues arise because of the unpredictable nonzero structure. Secondly, SpMV multiplication algorithms are inevitably memory-bound because the sparsity causes a low arithmetic intensity. Three state-of-the-art algorithms for parallel SpMV multiplication on shared-memory systems are discussed. Six new hybrid algorithms are developed which combine optimization techniques of the current algorithms. These techniques include parallelization strategies, storage formats, and nonzero orderings. A modern and high-performance implementation of all discussed algorithms is provided as open-source software. Using this implementation the algorithms are compared. Furthermore, SpMV multiplication algorithms require the matrix to be stored in a specific storage format. Therefore, the conversion time between these storage formats is also analyzed. Both tests are performed for multiple unstructured sparse matrices on different machines: two multi-CPU and two single-CPU architectures. We show that one of the newly developed algorithms outperforms the current state-of-the-art by $19\%$ on one of the multi-CPU architectures. When taking conversion time into consideration, we show that $472$ SpMV multiplications are needed to cover the cost of converting to a new storage format for one of the hybrid algorithms on a multi-CPU machine.
\end{abstract}

\begin{keywords}
    Sparse matrix-vector multiplication, shared-memory parallelization, unstructured sparse matrix, high-performance computing, sparse linear algebra, graph analysis
\end{keywords}

\begin{MSCcodes}
    68W10, 68--04
\end{MSCcodes}

\section{Introduction}
The sparse matrix-vector (SpMV) multiplication is an important kernel in many widely used algorithms for scientific computing and graph analysis. For example iterative methods for sparse linear systems such as GMRES, BICGSTAB, and QMR~\cite{saad}. Algorithms in the graph domain that rely on SpMV multiplication are breadth-first search and the famous PageRank algorithm~\cite{GAP-eval}. Because the SpMV multiplication is at the core of these, and many more algorithms, an efficient parallel implementation is key to fast execution. 

Two large sources of sparse matrices are discretizations of partial differential equations and graphs~\cite{SparseTamuEdu}. The discretizations are usually performed on a mesh and sparsity occurs because there are only local interactions. Furthermore, the local interactions make that the nonzero pattern of these matrices usually exhibits structure. As a result, most sparse matrices resulting from the discretization of a partial differential equation are structured matrices. Sparse problems with structured data can often be transformed into a problem that behaves similarly to a dense problem by exploiting the structure~\cite{SBD, Matrix-Partitioning}. A sparse matrix originating from a graph takes the form of an adjacency matrix. This is a matrix that has an entry for each edge between two nodes. Because most graphs do not originate from a 2D or 3D structure, the amount of edges per vertex usually varies a lot. For example, it has been observed that the amount of edges per vertex approximately follows a power law for some real-world graphs~\cite{PowerLawBiology, PowerLawInternet, PowerLawNature}. This large variation makes that the adjacency matrices of graphs are usually unstructured. In this work, we will focus on the second type of sparse matrices because this type of dataset is more frequent than ever due to the trend toward large-scale data collection. Furthermore, previous research is mostly focussed on structured sparse matrices. Therefore, we believe that an improvement in algorithm performance is still possible for unstructured sparse matrices.

Two main problems need to be taken into consideration when constructing an efficient parallel SpMV multiplication algorithm for unstructured sparse matrices. The first problem is caused by the sparsity of the input. It is not efficient to store a sparse matrix as a two-dimensional array because of the many zero elements. Therefore, specific storage formats are used. These only store the nonzero elements together with index information. Because of this, each arithmetic operation of the SpMV multiplication requires multiple memory operations, and thus the arithmetic intensity is low. On modern processors, this poses a problem because, while processor speed has increased dramatically, the memory speed remains slow in comparison~\cite{comporg-design}. To overcome this low arithmetic intensity, it is important to use the cache of modern processors as much as possible. The second problem is parallel load balancing. Unstructured sparse matrices have an unpredictable nonzero structure, so finding an adequate load balance is hard. To this end, strategies need to be found that split up the workload as efficiently as possible. However, the strategies themselves cannot take up too much time during multiplication.

This text reviews the current state-of-the-art algorithms for SpMV multiplication on shared-memory processors for unstructured matrices: Compressed Sparse Blocks, Row-Distributed Block CO-H and Merge-Based parallel SpMV multiplication. Moreover, it presents six new hybrid algorithms that combine different aspects of the current state-of-the-art. A modern and high-performance parallel \texttt{C++} implementation of all algorithms is provided as open-source software. As far as the authors are aware, a modern implementation is not available for two of the state-of-the-art methods. Furthermore, most high-performance algorithms require that the matrix is stored in a specific storage format. The conversion to these storage formats is also a time-consuming step. Therefore, the \texttt{C++} implementation also includes conversion methods. The performance of the SpMV multiplication and the conversion is thoroughly analyzed on different shared-memory machines. These include two uniform memory access (UMA) and two non-uniform memory access (NUMA) systems~\cite{comporg-design}. One could argue that all modern CPUs are NUMA architectures. However, in this work we will make the assumption that CPUs that only expose one NUMA domain to the operating system are UMA architectures. The tested UMA machines are single-CPU systems and the NUMA machines contain two CPUs.

When discussing algorithms in this work, the following notation will be used. The sparse matrix $A \in \mathbb{R}^{m \times n}$ with \texttt{nnz} nonzero elements, leads to the SpMV multiplication $y = Ax$ with dense vectors $x$ and $y$.

The remainder of this work is structured as follows. \Cref{sec:convStorage} introduces the basic storage formats for sparse matrices. These are the triplet or coordinate format, compressed row storage, and its variants. \Cref{sec:earlywork} discusses the current state-of-the-art algorithms for SpMV multiplication. Two of these algorithms introduce a new storage format that subdivides the sparse matrix into sparse subblocks. In \cref{sec:hybrid}, we present six new hybrid algorithms that combine aspects of various state-of-the-art algorithms. Some of these hybrid algorithms also introduce new storage formats. \Cref{sec:expDesign} first introduces the implementation of the SpMV multiplication algorithms and the conversion methods. Secondly, the test set containing multiple unstructured sparse matrices and the test machines are introduced. \Cref{sec:results} discusses the performance of the described SpMV multiplication algorithms. Furthermore, the conversion time is also measured and analyzed. Finally, \cref{sec:conclusion,,sec:futwork} present the conclusions and future work.

\section{Conventional sparse matrix storage formats}\label{sec:convStorage}
This section introduces four basic storage formats for sparse matrices. Furthermore, for each storage format the sequential SpMV multiplication algorithm is given. These basic storage formats and algorithms will form the basis for the state-of-the-art and hybrid SpMV multiplication algorithms.

The most straightforward way to store a sparse matrix is the triplet or coordinate format~\cite{sparse-matrix-chapter}. The matrix is stored using three arrays: \texttt{row\_ind}, \texttt{col\_ind}, and \texttt{data}, each with length \texttt{nnz}. The first two arrays store the row and column indices respectively and the last array stores the corresponding nonzero elements. The triplet format leads to an intuitive SpMV multiplication algorithm that loops over the arrays and performs
\begin{equation}
    y[\texttt{row\_ind}[i]] \mathrel{+}= \texttt{data}[i]*x[\texttt{col\_ind}[i]]
\end{equation}
in each iteration.

A more storage optimal scheme is Compressed Row Storage (CRS)~\cite{solution-algebraic-eig-problems}. For this storage format, the nonzero elements of the matrix need to be sorted in row-wise order. The \texttt{col\_ind} and \texttt{data} arrays are the same as for the triplet format. But, because of the specific ordering, the row indices can be compressed by only storing one element per row. This element points to the start of the row in the \texttt{col\_ind} and \texttt{data} arrays. The array that stores these row pointers, \texttt{row\_ptr}, will only have $m+1$ entries instead of \texttt{nnz}. \Cref{alg:SPMV-CRS} describes the SpMV multiplication algorithm for matrices stored using the CRS format. It proceeds by looping over the \texttt{row\_ptr} array and performing the multiplication row by row.

\input{CRS.tex}

The CRS format can be extended to the Incremental Compressed Row Storage (ICRS) format~\cite{ICRS-Koster}. The \texttt{col\_ind} array is replaced by a \texttt{col\_inc} array which does not hold column indices but positive increments between the subsequent column indices. Furthermore, row changes are signaled by letting the column index surpass $n$. The column index of the first element in the next column is then the calculated value modulo $n$. Because of this, the \texttt{row\_ptr} array can be replaced by a \texttt{row\_jump} array which stores the increments between subsequent row indices. The advantages of this format are twofold. First, because of the change to increments, pointer arithmetic can be used which is more efficient~\cite{ICRS-Koster}. Second, the change to row jumps means that if there are empty rows, the size of the \texttt{row\_jump} array will be smaller than $m$. 

Finally, the ICRS format can be generalized into the Bidirectional ICRS (BICRS) format~\cite{SpMV-Hilbert}. This format allows negative column and row increments. Overflow of the column index is still used to signal a row change. Because negative increments are possible, the nonzero elements can be ordered arbitrarily. This enables the use of specific nonzero orderings that increase the cache usage. However, an arbitrary nonzero ordering can lead to an increased storage cost if additional row jumps are necessary. 

The SpMV multiplication algorithm for the ICRS and BICRS formats is given in \cref{alg:SPMV-BICRS}. It is important to note that the first element of the \texttt{row\_jump} and \texttt{col\_inc} array correspond to the row and column index of the first nonzero element.

\input{BICRS.tex}

\section{State-of-the-art}\label{sec:earlywork}
This section discusses the current state-of-the-art parallel algorithms for SpMV multiplication. The work of Yzelman and Roose~\cite{high-level-SpMV} compares a large set of algorithms, and it shows that two methods have the best performance: Compressed Sparse Blocks (CSB) and Row-Distributed Block CO-H (BCOH). Later Merrill and Garland found that the Merge-Based SpMV algorithm has equal performance to CSB~\cite{SpMV-merge}.

\subsection{Compressed Sparse Blocks}
The Compressed Sparse Blocks~\cite{SpMV-CSB} (CSB) algorithm introduces a new storage format. In this CSB format, the matrix is subdivided into sparse blocks. Hence, there are two storage schemes for the matrix. One on the block level and one inside the blocks. The nonzero elements in the sparse subblocks are stored using the triplet format. Because the subblocks are smaller than the original matrix, the indices can be compressed. This compression is performed by packing the row and column index into one integer. The bits representing the integer are split in two and the row index takes up the first half of bits and the column indices the other half. These compressed indices are then stored in a single array where each subblock is stored contiguously. The nonzero entries are also stored in one contiguous array. On the block level, the matrix is stored as a dense matrix. The \texttt{blk\_ptr} array stores a pointer for each block in row-wise ordering. These point to the start of each subblock in the data and compressed indices array.

The motivation for sparse blocks is better cache reuse~\cite{SBD}. This is because for the multiplication of a single sparse block only a small part of the input and output vector is needed. Thus, there is a large probability that when an element of the input or output vector is used multiple times, it will remain cached during the block multiplication. Using a conventional sparse storage format, such as BICRS, this blocking would induce storage overhead due to the extra row jumps. However, because integer compression is used for the CSB storage format, the overhead is negligible compared to the CRS storage format if the block size is chosen such that the compressed row and column index fit inside the width of a standard integer~\cite{SpMV-CSB}.

For efficient SpMV multiplication, an appropriate subblock size needs to be used. Buluç et al.\ found~\cite{SpMV-CSB} that the optimal block size $\beta$ satisfies
\begin{equation}\label{eq:CSB-BlockSize}
    \lceil \log_2\left(\sqrt{n}\right)\rceil \leq \log_2(\beta) \leq 3 + \lceil \log_2\left(\sqrt{n}\right) \rceil.
\end{equation}
However, two other constraints need to be considered. The first one is storage cost. As described above, the overhead compared to CRS is negligible if the compressed indices can be packed in a $32$ bit integer. Thus, each index can maximally take up $16$ bits which means the block size $\beta$ cannot be larger than $2^{16}$. Secondly, for efficient cache utilization, the regions of $x$ and $y$ needed for multiplying the subblock should also fit comfortably in the L2 cache.

This leads to an implementation that sets $\log_2(\beta)$ to the upper bound of~\cref{eq:CSB-BlockSize} and checks the constraints. If a constraint is not met, $\log_2(\beta)$ is lowered by one until all constraints are satisfied.

The efficient SpMV multiplication algorithm corresponding to this storage format uses tasking to achieve parallel execution. The multiplication of each block row corresponds to one task, this is chosen to avoid false sharing. But, if a block row contains too many nonzero elements the task is split and temporary vectors are used to avoid false sharing. Finally, there is also the possibility to split up a dense block into multiple tasks. To perform this splitting efficiently, the nonzero elements in each subblock are stored in Z-Morton order~\cite{ZMorton}. This ordering first stores the elements of the top-left quadrant, then the top-right, bottom-left, and finally the bottom-right quadrant. This ordering is applied recursively in each quadrant. An example can be seen in~\cref{fig:ZMorton}. By using the Z-Morton order, dense blocks can be efficiently split using binary search. Furthermore, this ordering can also induce extra cache reuse inside blocks because the Z-Morton order provides additional locality.

\input{ZMortonRecursion.tex}

\subsection{Row-Distributed Block CO-H}
The Row-Distributed Block CO-H~\cite{high-level-SpMV} (BCOH) algorithm is also based on sparse subblocks. The nonzero elements in each subblock are stored using ICRS. As for the CSB storage format, the indices can also be compressed. However, because the ICRS format does not naturally group indices, the compression is not performed by packing the row and column indices into one integer, but two arrays with compressed size are kept. The storage of the sparse block matrix is also different from the CSB algorithm: no \texttt{blk\_ptr} array is used. Instead, the blocks are treated as elements of a sparse matrix and this sparse block matrix is stored using the BICRS storage format. The index arrays used to store the block matrix are also compressed. Thus, two compressed sparse storage formats are used: compressed BICRS on the block level and compressed ICRS inside each block.

A block size selection method was not mentioned in the state-of-the-art. Therefore, we decided that, as for CSB, each index increment of the BICRS and ICRS storage format can only take up a maximum of $16$ bits. Furthermore, the block size selection method used for CSB is also applied to this algorithm. However, the upper bound on $\beta$ had to be tweaked because the ICRS format used inside the blocks expects that overflow is possible. Thus, an upper bound of $2^{15}$ is set on the block size.

As for CSB, the motivation to use subblocks is better cache reuse. The authors believe that using a space-filling curve inside a subblock is unnecessary because the relevant part of the input and output vector will fit into the L2 cache~\cite{high-level-SpMV}. To improve the locality nonetheless, they induce an ordering on the subblocks themselves, which is possible because BICRS is used on the block level. The ordering which is chosen follows the Hilbert curve~\cite{Hilbert-Curve}. This space-filling curve differs from the Z-Morton curve because there are orientation changes involved. The recursive definition of the Hilbert curve is depicted in \cref{fig:Hilbert}.

\input{HilbertRecursion.tex}

Finally, to have an efficient parallel SpMV multiplication algorithm a parallelization strategy has to be chosen. For the BCOH algorithm, the load is balanced statically. This is performed by first dividing the rows of the matrix amongst the available threads such that each thread has approximately the same amount of nonzero elements. Then, each thread initializes its blocks by storing the relevant elements using the compressed ICRS storage format. Finally, the threads sort their blocks using the Hilbert curve and store the sparse block matrix using BICRS. The SpMV multiplication algorithm then proceeds with each thread executing the part of the multiplication corresponding to its rows in parallel.

\subsection{Merge-Based}
The Merge-Based parallel SpMV multiplication~\cite{SpMV-merge} is an algorithm that works directly on the CRS storage format. It uses an analogy to the parallel merging of a sorted list to try to obtain perfect load balancing.

The parallel merge-path algorithm used to merge two sorted lists in parallel, on which the Merge-Based SpMV multiplication algorithm is based, is explained first. This algorithm ensures that each thread performs the same amount of operations and thus consumes an equal amount of elements. An example execution of the algorithm is presented in \cref{fig:SPMV-MergePath}. A matrix represents the indices in both sorted lists. The algorithm starts from $(0, 0)$, with an empty merged list. In each step, the elements in both lists corresponding to the current indices are checked and the smallest is consumed. When an element is consumed the corresponding index is increased by one. When the algorithm reaches the end of both lists, the bottom right of the matrix, the execution finishes and a sorted merged list is formed. To parallelize this algorithm, the matrix corresponding to the indices is subdivided amongst threads. This is done by dividing the matrix into bands where each thread holds an equal amount of diagonals. At the start of the algorithm, each thread calculates its beginning and endpoint as the point $(i,j)$ on the respective diagonal where $A[i] \leq B[j]$ but also $A[i+1] > B[j]$ with $A$ and $B$ representing the sorted lists. Using this start and end point each thread performs their respective part of the merging of both lists. Because the bands assigned to each thread have the same amount of diagonals, each thread consumes the same amount of elements. An example can be seen in \cref{fig:SPMV-MergePath} where each thread consumes four out of the twelve elements.

\input{merge.tex}

\footnotetext{This figure is an adaptation of Figure 7 from the work of Merrill and Garland~\cite{SpMV-merge}.}

This algorithm is adapted to a CRS SpMV multiplication algorithm as follows. The \texttt{row\_ptr} array without the starting zero is taken as the first list and the natural numbers, corresponding to the indices in the {col\_ind} and \texttt{data} arrays, as the second list. The algorithm starts by initializing a temporary value \texttt{temp} to zero. Then, the algorithm is executed as described above, but consuming an element triggers an action instead of creating a merged list. Consuming an element $k$ of the natural numbers induces the multiply-add operation
\begin{equation}
    \texttt{temp} = \texttt{temp} + \texttt{data}[k] * x[\texttt{col\_ind}[k]],
\end{equation}
which corresponds to line 4 and 5 of \cref{alg:SPMV-CRS}. If an element of the \texttt{row\_ptr} array is consumed the end of a row is signaled. The temporary value is output to $y$ and set to zero for the next row.

The parallelization of the SpMV multiplication algorithm is performed as for the original sorting algorithm. However, when the final element for a thread is consumed it is possible that the temporary value is not output to $y$. Therefore, each thread stores its last value of \texttt{temp} and the corresponding row index. When all threads have finished, these stored values are sequentially added into $y$. Another advantage of this is that during the parallel stage of the execution, each thread accesses a distinct part of $y$. Thus, false sharing is avoided as much as possible. As for the merge-path algorithm, this algorithm also ensures perfect load balancing in the sense that each thread executes the same amount of operations where an operation is a multiply-add operation or storing the temporary value in the output vector.

\section{Hybrid algorithms}\label{sec:hybrid}
The algorithms discussed in the previous section use three techniques to obtain faster execution time for SpMV multiplication: blocking, sorting nonzero elements, and load balancing. Blocking and sorting are used to obtain better locality and thus better cache reuse. An appropriate load balancing is needed to achieve efficient parallelization for arbitrary inputs. There is a wide range of possibilities for static and dynamic scheduling.

Because of the large pool of performance-enhancing techniques, the algorithms described in the previous section usually take a specific route that the authors deemed best. But, because all state-of-the-art algorithms obtain high performance for SpMV multiplication it is also interesting to combine different aspects of multiple algorithms. Therefore, six new \textit{hybrid} algorithms and their corresponding storage format will be explored in this section.

\subsection{Compressed Sparse Blocks}\label{sec:CSBH}
A new algorithm was created by altering the CSB storage format. The ordering inside blocks was changed from the Z-Morton curve to the Hilbert curve. The new SpMV multiplication algorithm will be called the Compressed Sparse Blocks Hilbert (CSBH) algorithm.

This \textit{hybrid} algorithm is inspired by the observation that the locality provided by the Z-Morton ordering is suboptimal compared to the Hilbert curve. This can easily be seen by comparing~\cref{fig:ZMorton,,fig:Hilbert}. The Z-Morton curve induces large jumps across diagonals. For example, when moving from the top-right to the bottom-left quadrant the row index increases by one but the column index decreases by the number of columns of the matrix. This never happens for the Hilbert curve, where only one index changes at a time and the index value always increases or decreases by one. However, if we sort the nonzero elements of a matrix using these curves, the nonzero structure also has to be considered. Therefore, both orderings will induce sudden jumps in row and column index. Nonetheless, we will show that the Hilbert curve provides better locality.

The only algorithmic change needed to support the new storage format is the dense block splitting. A binary search method can still be used to find the start and end point of each subblock, but the different ordering needs to be taken into account. Furthermore, the orientation of the Hilbert curve changes on different recursion levels~\cite{Hilbert-Curve}. This can be seen in \cref{fig:Hilbert}, where the orientation changes based on the quadrant of the submatrix in which the Hilbert curve is applied recursively. Thus, to correctly find the splitting points for the nonzero elements of dense blocks, we need to keep track of the orientation of the Hilbert curve when applying this splitting recursively. This induces more complexity than the original sorting method. However, the performance impact will be negligible because the majority of the blocks are sparse and most time is spent multiplying the nonzero elements.

\subsection{Row-Distributed Block CO-H}\label{sec:Hybrid-BCOH}
Subsequent alterations to the BCOH algorithm lead to three new \textit{hybrid} algorithms. The first alteration consists of changing the storage format inside the blocks. The ICRS storage format is changed to a compressed triplet storage format. Furthermore, the compressed indices are, as for CSB, stored by packing them into a $32$ bit integer. The ordering of nonzero elements remains row-wise. The described alteration to the BCOH algorithm is called the BCOH Compression (BCOHC) algorithm.

The original BCOH algorithm uses compressed ICRS inside each block. The reason for this is that the storage cost of a CRS-based format is lower than for the triplet format~\cite{sparse-matrix-chapter}. But, this scheme also has its downsides. Firstly, the ICRS storage format does not allow arbitrary nonzero ordering. Second, hardware simulations have shown that the average DRAM bandwidth usage of the BCOH algorithm is significantly lower than for other SpMV multiplication algorithms~\cite{Kobe-Thesis}. Because most time is spent in the block multiplication step, the memory usage pattern can be attributed to the storage format used inside the blocks.

Both of these concerns are addressed by changing to the block storage format used in CSB: compressed triplet format. This compression makes an arbitrary nonzero ordering possible, and it does not suffer from suboptimal DRAM usage. The downside of changing from ICRS to the triplet format is that the storage cost increases because for each nonzero element, the indices take up $32$ bits of space. This is not the case for compressed ICRS, where each element takes $16$ bits for the column index and each nonempty row adds another $16$ bits of storage.

The second \textit{hybrid} algorithm based on BCOH is the BCOH Compression Hilbert (BCOHCH) algorithm. This algorithm uses the storage format of BCOHC but sorts all nonzero elements per thread using the Hilbert curve. So, the rows of the sparse matrix are first divided amongst the threads, as is done for the original BCOH algorithm. Then, all nonzero elements per thread are sorted in Hilbert order. Finally, the elements are stored inside their respective block using the compressed triplet format. Because of the recursive definition of the Hilbert curve, the sparse block matrix will still be sorted in Hilbert order per thread.

Sorting all nonzero elements per thread using the Hilbert curve has two advantages. First, the elements inside each block are ordered following the Hilbert curve. For most matrices, this will increase locality and lead to better cache reuse. Secondly, because for each thread all elements are sorted using the Hilbert curve, the transitions between blocks are also as local as possible. This can be seen by interpreting the blue curve of the left part of \cref{fig:Hilbert} as a two-by-two block matrix with blocks of size two-by-two. We can then easily see that during the transition from one block to the next, the row or column index always increases by one.

Finally, a last algorithmic change was explored. The BICRS storage format on the block level seemed wasteful for denser matrices because almost all blocks will be filled for those types of matrices. Thus, the BCOHCH algorithm was changed to use a \texttt{blk\_ptr} array as is used in the CSB storage format. However, this array does not point to the blocks in row-wise ordering but in Hilbert order. The algorithm that uses dense block storage is called BCOH Compression Hilbert Pointer (BCOHCHP).

The major advantage of this scheme is apparent: if the block matrix is almost dense, the storage cost can be reduced significantly. The BICRS storage format on the block level stores three arrays: \texttt{block\_nnz}, \texttt{row\_jump\_block}, and \texttt{col\_jump\_block}. The first one stores $32$ bit integers and the others store $16$ bit integers. If we assume a dense block matrix, we can estimate the size of the row jump array to be half the number of blocks. This is because when following the Hilbert curve we can only have a row or a column jump, and they occur mostly evenly throughout the curve. The column jump and block nonzero count arrays have an entry for every block. The dense storage on the block level stores the \texttt{block\_ptr} array which has a size equal to the number of blocks and each element is a $32$ bit integer. From this, it is clear that the dense storage at the block level decreases the storage cost by approximately $1.5$ $16$ bit integers per block. For the whole matrix, dense storage takes about $4/7$ of the storage space required by the BICRS storage format.

However, the major downside of this change is that because the \texttt{blk\_ptr} array directly points to the Hilbert order, calculations are needed to compute the row and column indices of each block. This is not needed for the original algorithm because the BICRS storage format indicates the block row and block column increments. In essence, the BCOHCHP algorithm trades storage space for extra calculations.

\subsection{Merge-Based}
The description of the Merge-Based SpMV multiplication algorithm sounds promising as it provides perfect load balancing if the described operations both take the same time. One of the downsides of this algorithm is that it works directly on the CRS format and thus the matrix must always be traversed row-wise. This means that, for most unstructured matrices, the cache efficiency will not be great. Therefore, we created two new algorithms that adapt the original algorithm to use blocking. The first adaptation is called Merge Blocking (MergeB).

To be able to use the Merge-Based SpMV multiplication algorithm we subdivide the matrix into sparse blocks and the blocks themselves act as nonzero elements of a sparse block matrix. This sparse block matrix is ordered row-wise and stored using the CRS storage format. The data of this CRS structure is a pointer array that points to the start of each block. Inside the blocks, the elements are stored using the compressed triplet format and remain sorted in row-wise order. As described in \cref{sec:Hybrid-BCOH}, this format is superior in terms of DRAM bandwidth utilization. The block size is chosen in the same manner as for the CSB algorithm. This allows for an easy comparison between all algorithms.

Because the CRS structure is still present on the block level, we can use the Merge SpMV multiplication algorithm with some adaptations. The multiply-add operation is replaced with a sparse block multiplication and the temporary value is replaced with a temporary vector. The output operation must then also be replaced by copying the temporary vector into the correct part of the output vector.

As seen in the previous sections, a space-filling curve can be used to increase the locality and thus the cache efficiency. For the MergeB algorithm, a space-filling curve can only be used inside the sparse blocks because the CRS storage format must be used on the block level and this storage format does not allow for arbitrary ordering. So, for the second adaptation, the nonzero elements inside each sparse block are sorted using the Hilbert curve. The Hilbert curve was chosen because it provides better locality properties than the Z-Morton curve as explained in \cref{sec:CSBH}. This algorithm is called Merge Blocking Hilbert (MergeBH).

\section{Experimental design}\label{sec:expDesign}
This section describes the evaluation method of the nine algorithms described above. We not only benchmarked the SpMV multiplication algorithms, but also the time needed to convert from the triplet format to the specific storage format each algorithm requires. To enable a comprehensive analysis of the runtime, a high-performance implementation and a set of test matrices is introduced. All algorithms are tested on four different machines, of which two are NUMA.

\subsection{Implementation}
An efficient implementation of the algorithms described in \cref{sec:earlywork} and \cref{sec:hybrid} is provided as open source software\footnote{\url{https://gitlab.kuleuven.be/numa/public/spmat_public/tree/v1.0}}. The software is written in \texttt{C++} and all tested methods are parallelized using \textit{OpenMP}~\cite{openmp5}. The code is compiled using the \texttt{dpcpp} compiler, which is part of the \textit{Intel oneAPI} programming model~\cite{oneapi}. Compilation is performed using the \texttt{-DNDEBUG},  \texttt{-O3}, and \texttt{-march=native} flags. Before execution, the \textit{OpenMP} environment variable \texttt{OMP\_PROC\_BIND} was set to true to bind threads to cores.

To enable an in-depth analysis of the SpMV multiplication algorithms, two additional baseline implementations are tested: The sequential CRS multiplication algorithm described in \cref{alg:SPMV-CRS} and a parallelized version of this algorithm. \textit{OpenMP} is used to parallelize the outer loop (line 2). Dynamic scheduling with a chunk size of $512$ was found to provide the best performance. The parallelization will be called ParCRS for the remainder of this work.

Some changes were made to have an efficient execution of the SpMV multiplication algorithms on NUMA architectures. On machines with multiple NUMA domains, pieces of data are always stored closest to one domain. This makes that the cores of that domain will have a significantly faster access time compared to the other cores. Because of this, the operating system will try to obtain an optimal data allocation by moving data between sockets during execution~\cite{DynamicPageMigration}. This hurts performance because these intersocket data movements are slow~\cite{NUMA}. However, even though intersocket data movement is inevitable, it can be minimized using interleaved allocations. These are allocations where chunks of data are interleaved between the different memory banks. This means that some data will always be present in the closest memory bank. On NUMA architectures, interleaved allocation is always used for the input vector because it will be accessed from all CPUs. For the output vector and matrix storage, the interleaved allocation is only used if the parallelization is performed dynamically (ParCRS and CSB).

A parallel implementation of the conversion between the triplet storage format and the storage formats required by each of the SpMV multiplication algorithms is also provided. The conversion is different for each storage format, but always consists of two steps. First, the nonzero elements are sorted in the specific ordering needed for the storage format. For example, for the BCOHCH storage format, the nonzero elements are sorted based on two criteria: thread and Hilbert order. Secondly, the different arrays for the storage format are populated by copying, and potentially compressing, the nonzero elements and index information. The first step is usually the most costly because it has been proven that the best average complexity of a sorting algorithm is $\mathcal{O}(n \log{n})$~\cite{IntroToAlgorithms}. The second step only accesses every element once.

For the sorting step we use the In-place Parallel Super Scalar Samplesort algorithm~\cite{IPS4o}. This parallel sorting algorithm performs the sorting in-place and also provides a high cache-efficiency. The original source code for this algorithm could directly be used by implementing a custom iterator that loops over the index and data arrays of the triplet format simultaneously. To enable sorting for the specific storage formats, custom comparison operators are implemented.

\subsection{Test setup}\label{sec:testSetup}
To make the tests as representable as possible, the methods are tested on multiple unstructured square sparse matrices. The test set is presented in \cref{tab:Matrices}. All matrices are obtained from the Florida sparse matrix collection~\cite{SparseTamuEdu}, except for \texttt{HHH} and \texttt{LHH} which are randomly generated. The test matrices were selected such that a wide range of possible parameters is covered. We mainly considered the density and size of the matrix. The density is calculated as the amount of nonzero entries to the total amount of entries. We split up the test set in two classes: low density matrices with a density lower than $10^{-6}$, and matrices with a density higher than this threshold. Furthermore, we also made sure that the matrices were large enough such that the timings are reliable. The fastest execution time of the SpMV multiplication algorithms for the smallest matrix (\texttt{HHH}) is in the order of a few milliseconds.

\input{TestMatrices.tex}

Another metric that is important for the execution time of the SpMV multiplication and the storage format conversion is the machine on which the tests are performed. Important parameters are the amount of CPU cores and their speed. Furthermore, the main memory also significantly affects performance as the SpMV multiplication is inevitably memory-bound. Finally, the amount of CPUs will also have an effect. If multiple CPUs are present, communication across CPUs will impact performance. 

Because of the above, we tested the different algorithms on four machines. Their hardware characteristics are summarized in \cref{tab:Machines}. Each machine is named after the microarchitecture of the CPU. The Sapphire Rapids machine has two Intel Xeon Platinum 8468 CPUs. This machine is configured with 4 NUMA domains per CPU. Ice Lake NUMA consists of two Intel Xeon Platinum 8360Y CPUs. Each CPU is configured with one NUMA domain. Ice Lake UMA is the same machine, but only one CPU is used to avoid NUMA effects. Finally, the Cascade Lake machine has one Intel Core i9--10980XE CPU with hyperthreading enabled. This CPU is also configured with one NUMA domain.

\input{TestMachines.tex}

\footnotetext{For conciseness the names of the following matrices are shortened: \texttt{mawi\_201512020130}, \texttt{hugebubbles-00020}, \texttt{hugetrace-00020}, \texttt{soc-LiveJournal1}, \texttt{kron\_g500-logn21}.}

The different SpMV multiplication algorithms were tested on all matrices with different amounts of threads. For the Sapphire Rapids machine, these ranged from $52$ to $96$ with increments of $4$. Ice Lake NUMA was tested using $40$ to $72$ threads in increments of $4$. For the UMA tests on this machine the amount of threads ranged from $16$ to $36$ in increments of $2$. Cascade Lake has fewer physical cores, thus the number of threads started at $6$ and went up until $36$ in increments of $2$. Each test consisted of $550$ timed executions. The result is the minimum measured timing. Finally, these timings are compared to the sequential CRS algorithm, and the parallel speedup is calculated.

Transforming a matrix to a storage format needed for a specific algorithm can be a costly operation. The cost of this operation depends on the complexity of the storage format and the required ordering of the nonzero elements. Because of this, the conversion step was also benchmarked. These tests were run on all machines for all test matrices with the same thread counts as reported above. The test consisted of $25$ timed executions. The results are presented as minimum conversion time divided by the fastest execution of the ParCRS SpMV multiplication. We chose to present the conversion time with respect to the ParCRS algorithm because this is the simplest algorithm based on the most common storage format CRS.

\section{Results}\label{sec:results}
We describe the results of the tests introduced in the previous section. First, we discuss the measured parallel speedup for the SpMV multiplication algorithms. Secondly, the conversion time between the triplet storage format and the specific storage formats needed for the SpMV multiplication algorithms is presented.

\subsection{SpMV multiplication}
The results of the tests of the different SpMV multiplication algorithms are summarized in \cref{tab:AggregateResultsL} and \cref{tab:AggregateResultsH}. The first table displays the mean parallel speedup compared to the sequential CRS algorithm for the test matrices with a density smaller than $10^{-6}$. The second table aggregates the results for the more dense test matrices. The results of the \texttt{mawi\_0130} matrix are left out of the first table. This is because this matrix has one row which is almost dense, and this makes that some algorithms perform significantly worse compared to their performance on other types of matrices. The results for \texttt{mawi\_0130} are discussed separately at the end of this section.

\input{AggregateL.tex}
\input{AggregateH.tex}

By looking at both tables we see that for the denser matrices, a larger speedup can be obtained. This effect can be attributed to better cache reuse and is explained by comparing two matrices with different densities and an equal number of nonzero entries. The amount of FLOPS for an SpMV multiplication will be the same for both matrices because they have the same amount of nonzero entries. However, because the matrix with higher density will be smaller, the input and output vector will also be smaller. Therefore, the chance for cache reuse is higher for denser matrices because fewer distinct elements of $x$ and $y$ need to be accessed. This means that the memory bottleneck will be smaller, and the parallel speedup can become larger.

From the results, the importance of RAM speed is also immediately clear. The Cascade Lake machine has a significantly lower RAM speed than the other machines. This is noticeable in all results, where the achieved speedup is less than one-third of that on the other machines for all algorithms. It is clear that this effect can not only be attributed to the smaller core count of the Cascade Lake CPU. Furthermore, the low RAM speed hinders extra speedup when using additional threads. For matrices with high sparsity, each algorithm achieves peak performance when using $12$ threads or fewer. This shows that memory is a more significant bottleneck than computing power for all algorithms.

The performance of the algorithms based on the CRS storage format (ParCRS and Merge) is usually pretty equal. The Merge algorithm only performs significantly better for the matrices with lower density on the Ice Lake NUMA machine and for the matrices with higher density on both NUMA machines. This indicates that the load balancing is quite equal between both algorithms. The Merge algorithm has an advantage on NUMA systems because static scheduling results in less intersocket data movement. Compared to the other algorithms, we see that the ParCRS and Merge algorithms almost never significantly outperform the other algorithms, except for the Merge algorithm on the Ice Lake NUMA machine for matrices with high sparsity.

The CSB variants have the best performance for matrices with low density on the UMA machines. We assume this is only the case for these machines because dynamic scheduling does not perform that well on NUMA machines because of intersocket data movements. Furthermore, the performance gap compared to the other algorithms is higher for the Cascade Lake machine than for Ice Lake UMA. This is probably because dynamic scheduling performs better if the RAM speed is a limiting factor. In that case, the time taken to process a chunk of work is more unpredictable due to memory queueing delays. For matrices with higher density, the CSB variants are always outperformed. Finally, the Hilbert ordering used for CSBH has a positive effect on most machines. It never has a significantly negative effect.

The BCOH algorithm or one of its variants is the fastest algorithm on the NUMA machines for matrices with higher density. For the UMA machines, performance close to that of the best performing algorithms is measured. For matrices with higher sparsity, the BCOH variants are the best performing on the Sapphire Rapids machine, perform a bit worse than the best algorithm on the Ice Lake NUMA machine, and significantly worse on the UMA machines. This difference can probably again be attributed to the avoidance of intersocket data movement on NUMA machines because of the static scheduling. This explains why this algorithm works best on the Sapphire Rapids machine, as it has eight NUMA domains compared to the two NUMA domains of the Ice Lake NUMA machine. Distinguishing between the variants we see that changing the compression scheme has a significant positive impact on the algorithm, especially for the denser matrices. Only on Cascade Lake, the speedup decreases a little. Additionally, sorting the nonzero elements inside each block almost always increases the performance, as discussed for CSBH. Finally, changing from BICRS on the block level to block pointers always has a negative effect. This is especially noticeable for matrices with low density. This is logical because these matrices will have a large amount of empty blocks, which can lead to a higher storage cost of pointers compared to BICRS.

Finally, we inspect the blocking variants of the Merge algorithm. These methods are only relevant on the Cascade Lake machine. On this machine we see that the algorithm performs almost equally well to the best algorithm for matrices with a low nonzero count and obtains optimal performance for more dense matrices. On the other machines there is always a significant performance gap with the other methods. We suspect that this is because on the block level, the matrix has dimensions in the range of a couple of thousand rows and columns. Therefore, the Merge algorithm executed on this block matrix will have not be able to provide the necessary load balancing because there are very few blocks. We suspect that the reason this algorithm works well on the Cascade Lake machine is because the core count is low and thus the blocks to thread ratio is higher on this machine. As for the other algorithms, the Hilbert ordering inside blocks only has a positive effect.

The results for the \texttt{mawi\_0130} matrix are presented in \cref{tab:MawiSpeedup}. From these results it is immediately clear why this matrix is presented separately. Because this matrix contains one row which is almost dense, only the algorithms which allow multiple threads to process a single row obtain a good speedup: CSB, CSBH, and Merge. The Merge Blocking variants can also split up a row, but the load balancing of these algorithms is not based on the amount of nonzero elements per block. So, these variants will also not perform well on this type of dataset. If we compare the best speedup for the \texttt{mawi\_0130} matrix to the best speedup reported for the other test matrices with low density (\cref{tab:AggregateResultsL}), we see that this type of dataset remains challenging even though rows can be processed by multiple threads simultaneously. This is because writing to the same output element from multiple threads introduces significant overhead to avoid read-write conflicts.

\input{MawiSpeedup.tex}

\Cref{fig:AvgParSpeedup} provides another comparison of the different algorithms. The average parallel speedup for all test matrices except the \texttt{mawi\_0130} matrix, is plotted against the amount of threads used. From this figure, we notice that the relative difference between the algorithms does not significantly change when changing the number of threads. Therefore, if an algorithm is fastest for a specific number of threads it will usually be the fastest for all possible amounts of threads. Secondly, in \Cref{fig:AvgParSpeedupEclipse}, we see that hyperthreading harms the parallel speedup on the Cascade Lake machine. When more than $18$ threads are used the speedup does not increase. Furthermore, for the algorithms with static scheduling, a clear performance decrease can be observed when comparing the results for $18$ threads to the results for $20$ threads. This can be explained by examining the CPU thread scheduling. The operating system tries to schedule as many threads on separate cores as possible. When running the program with $18$ threads, each core executes one thread. But, when $20$ threads are used, two cores will execute two threads. The threads that need to share a core will inevitably be slower than the others. This significantly slows down algorithms that use static scheduling because they assume that each thread is equally fast.

\input{AggregateSpeedup.tex}
\subsection{Storage format conversion}

\cref{tab:ConversionL} and \cref{tab:ConversionH} display the time needed to convert a matrix with respectively low and high density of nonzero elements to a specific storage format. The time needed is displayed as the number of SpMV multiplications of the ParCRS algorithm. In the table for the low density matrices, the results for \texttt{mawi\_0130} are again left out.

\input{ConversionL.tex}
\input{ConversionH.tex}

If we compare the tables in general, it is clear that the relative conversion cost is always higher for a denser matrix. This can be explained by noting that the conversion time of a really sparse matrix will not differ that much from a matrix with higher density. The only difference in conversion between these two types of matrices is that for the denser matrix, the data structures can be smaller. For example, the \texttt{row\_ptr} array in a CRS storage format will usually be smaller when the density increases and the amount of nonzero elements remains equal, because this makes that the amount of rows usually decreases. However, the difference in conversion time will be negligible. Therefore, the difference between the relative conversion time for really sparse matrices and more dense matrices is explained by the fact that the SpMV multiplication is faster for more dense matrices, as seen in the previous section.

Another interesting observation is that the cost of converting to Merge Blocking is almost always lower than converting to the CRS format. This is only not true on the Cascade Lake machine for matrices with high sparsity. When comparing both conversions we notice that the sorting step is a bit more involved for the Merge Blocking storage format because the nonzero elements need to be sorted in row-wise order on the block level and inside the blocks. So, the sorting step will be more expensive for the Merge Blocking storage format. In the second step of the conversion process, the Merge Blocking storage format has an advantage over CRS because it uses the triplet format inside blocks. Therefore, no extra calculations are needed, the nonzero elements only need to be copied, and the indices need to be compressed into 16 bit block indices. For the CRS storage format extra calculations are needed to compress the row indices into the \texttt{row\_ptr} array.

Finally, it is also interesting to inspect the different nonzero ordering strategies inside the blocks. Inside the blocks three types of ordering are used: row-wise, Z-Morton and Hilbert. To get a sense of the cost of Z-Morton ordering we can look at the difference in conversion time between the CSB storage format and the Merge Blocking storage format. These store the matrix differently on the block level but inside the blocks the storage format is equal. Hence, we can compare them because the calculations on the block level are negligible compared to the calculations inside the blocks. This comparison shows that the cost of converting to CSB is maximally $70\%$ higher than the cost of converting to Merge Blocking. The cost of Hilbert ordering is significantly higher. This can be seen by comparing the Merge Blocking  storage format and its Hilbert ordered variant, as the only change made is that Hilbert ordering is applied inside the blocks. Here we see that in the worse case the cost of converting a matrix can increase by a factor of $14$. When changing to Hilbert ordering for the BCOHC and CSB storage formats, this cost is lower. This is because the BCOHC already imposes a Hilbert order on the blocks and for CSB the Z-Morton order is used which is more expensive than row-wise ordering. 

\section{Conclusion}\label{sec:conclusion}
In this text, we first introduced the current state-of-the-art parallel algorithms for SpMV multiplication on shared-memory processors: one algorithm based on the standard CRS storage format and two algorithms that introduce a new storage format based on sparse blocking. Six new hybrid algorithms were developed by combining aspects of the different state-of-the-art algorithms. The methods used by the algorithms to increase the speed of the SpMV multiplication include load balancing techniques, blocking, and specific sorting of nonzero elements.

The current state-of-the-art and new algorithms were tested on a set of sixteen unbalanced square test matrices. These tests were run on four different machines using different amounts of threads. Two of these machines were NUMA. Furthermore, the time needed to convert from the triplet format to the specific storage format used by the different algorithms was also measured.

From tests of the SpMV multiplication algorithms, we concluded that all algorithms achieve a larger parallel speedup for denser sparse matrices because these matrices enable better cache reuse. Furthermore, the RAM speed is a limiting factor in the maximum parallel speedup that can be obtained and the amount of cores needed to achieve this. Additionally, it is also important to check the properties of the matrix before choosing an SpMV multiplication algorithm. For example, one test matrix contained an almost dense row. This made that only the Merge, CSB and CSBH algorithm obtained a relevant parallel speedup for this matrix because they can process a single row using multiple threads. Lastly, hyperthreading can hurt the performance of all algorithms, especially those that use static scheduling.

Looking more in-depth at the algorithms, the results show that the performance depends on the density of the matrix and the type of the machine. For matrices with low density, we see that on NUMA systems the algorithms based on the standard CRS storage format and the BCOHC or its Hilbert-sorted variant perform the best. On systems with one CPU, the CSB algorithm and its Hilbert-sorted variant performs best. For more dense matrices, the BCOHC and its Hilbert sorted variant performs best on almost all systems. On the UMA system with high RAM speed, the CRS based algorithms achieve equal performance to the BCOHC algorithms. On the UMA system with low RAM speed, the CRS variants clearly outperform the BCOHC algorithms. Furthermore, the Merge Blocking algorithm and its Hilbert-sorted variant achieve equal performance to the algorithms based on the CRS storage format.

From these results, we can conclude that the newly developed BCOHC, and its Hilbert-sorted variant, provide excellent results for higher-density matrices on NUMA systems. For these situations, the average parallel speedup of the BCOHCH algorithm can be up to $19\%$ higher compared to the current state-of-the-art. For matrices with lower density we see that both algorithms outperform the current state-of-the-art on a machine with high core count and many NUMA domains, and perform almost equally to the state-of-the-art on the other NUMA machine. Thus, the BCOHC and BCOHCH algorithms provide a good baseline algorithm to use on NUMA systems when processing matrices with different characteristics.

For UMA systems the results are more clear. The CSB and CSBH algorithms clearly outperform the other algorithms for matrices with low density. For matrices with higher density of nonzero elements, the algorithms based on the CRS storage format achieve the highest parallel speedup.

The tests that measured the conversion time between the triplet format and the other storage formats revealed that this is also an important metric because the cost of conversion can be high. In general, the conversion to the Merge Blocking storage format is the cheapest. Converting to the CRS and CSB storage formats is a bit more expensive. The conversion time for the BCOH and BCOHC storage formats has a high variability. The cost is between two and nine times higher than converting to CRS. Finally, for the blocking storage formats, sorting the elements inside blocks in the Hilbert order can increase the conversion cost by a factor of $14$ for the Merge Blocking algorithm. The cost for the BCOHC and CSB storage formats is lower but still remains high.

These results put the speed of the SpMV multiplication algorithms in perspective. For example, we see that sorting the elements in the Hilbert order always has a positive or negligible effect for the blocking algorithms. Based on these results one would be eager to always use these algorithms. However, when looking at the conversion time, the extra increase in performance is usually only worth it if many SpMV multiplications are performed for the same matrix. 

The amount of SpMV multiplications that need to be performed to justify the extra cost of converting to a new storage format can be calculated for the SpMV multiplication algorithms that perform better than the standard CRS algorithm. The BCOHC algorithm on the Sapphire Rapids machine requires $472$ SpMV multiplications on average to obtain a performance benefit. For BCOHCH this is $1525$ because the Hilbert ordering is more expensive. For CSB on a UMA machine with low RAM speed only $26$ SpMV multiplications are needed, and for CSBH $420$ multiplications are necessary to be more performant than the CRS storage format.

The thorough analysis of multiple high-performance algorithms shows that choosing an algorithm for SpMV multiplication is not straightforward. Different machine and matrix properties call for different algorithms to be used. Furthermore, the amount of times a specific matrix will be used in an SpMV multiplication is also of interest. If many SpMV multiplications are required we would recommend the BCOHC or BCOHCH algorithms on NUMA machines. On UMA machines the Merge algorithm seems to have a low conversion cost and good performance. If many SpMV multiplications are performed using the same matrix on UMA machines and the matrix is really sparse, the CSB or CSBH algorithms will perform even better.

\section{Future Work}\label{sec:futwork}
In this paper, hybrid algorithms were developed by combining aspects of different state-of-the-art algorithms. We did not consider low-level optimizations. These include: tuning block size, compressed storage formats, specific sorting techniques, vectorization, etc. We suspect that these optimizations can still provide a relevant speedup, but they will be largely machine-specific because they rely on cache, memory, and CPU properties. To better deal with this, it would be interesting to look into automatically tuning these parameters, like performed in the pOSKI library~\cite{pOSKI}.

Investigating the effect of RAM speed on the execution time can also lead to interesting results. As seen in the results for the machine with one CPU and low RAM speed, increasing the thread count might lead to worse results. Thus, it could be interesting to create an analytical model that can predict the most optimal amount of threads for a given machine's characteristics.

Finally, to use a state-of-the-art SpMV multiplication algorithm a conversion from the triplet format is almost always necessary. Therefore, it would be interesting to look into new ways of storing a sparse matrix on disk that keeps in mind the storage formats used for the state-of-the-art algorithms. The cost of disk space is ever decreasing~\cite{DiskPrice}, so it seems that there are possibilities to create a new format that stores multiple copies of the sparse matrix with different nonzero orderings.

\section*{Acknowledgments}
The research of Raf Vandebril was partially supported by the Research Council KU Leuven (Belgium), project C16/21/002, and by the Fund for Scientific Research --- Flanders (Belgium), projects G0A9923N and G0B0123N. The work of Karl Meerbergen is supported by the KU Leuven Research Council (Belgium), project C24E/19, and by the Fund for Scientific Research --- Flanders (Belgium), projects G088622N and G004124N. Some resources and services used in this work were provided by the VSC (Flemish Supercomputer Center), funded by the Research Foundation --- Flanders (Belgium) and the Flemish Government.

\bibliographystyle{siamplain}
\bibliography{references}

\end{document}

%% file: CRS.tex
\begin{algorithm}
    \caption{SpMV multiplication for the CRS format}\label{alg:SPMV-CRS}

    \begin{algorithmic}[1]
        \STATE{\textbf{Input:} \texttt{col\_ind}, \texttt{row\_ptr} and \texttt{data} arrays corresponding to $A \in \mathbb{R}^{m \times n}$. Dense input vector $x$.} 
        \STATE{\textbf{Output:} Dense output vector $y$ such that $y = Ax$.}

        \STATE{$y = 0$}
        \FOR{$i = 0$; $i < m$; $i \mathrel{+}= 1$}
            \FOR{$k = \texttt{row\_ptr}[i]$; $k < \texttt{row\_ptr}[i+1]$; $k \mathrel{+}= 1$}
                \STATE{$j = \texttt{col\_ind}[k]$}
                \STATE{$y[i] = y[i] + \texttt{data}[k]*x[j]$}
            \ENDFOR%
        \ENDFOR%
    \end{algorithmic}
\end{algorithm}

%% file: BICRS.tex
\begin{algorithm}
    \caption{SpMV multiplication for the (B)ICRS format}\label{alg:SPMV-BICRS}

    \begin{algorithmic}[1]
        \STATE{\textbf{Input:} \texttt{col\_inc}, \texttt{row\_jump} and \texttt{data} arrays corresponding to $A \in \mathbb{R}^{m \times n}$. Dense input vector $x$.} 
        \STATE{\textbf{Output:} Dense output vector $y$ such that $y = Ax$.}

        \STATE{$y = 0$}
        \STATE{$k = 0$; $r = 0$; $j = \texttt{col\_inc}[0]$; $i = \texttt{row\_jump}[0]$}
        \WHILE{$k < \texttt{nnz}$}
            \WHILE{$j < n$}
                \STATE{$y[i] = y[i] + \texttt{data}[k]*x[j]$}
                \STATE{$j = j + \texttt{col\_inc}[k]$; $k = k + 1$}
            \ENDWHILE%
            \STATE{$j = j - n$; $i = i + \texttt{row\_jump}[r]$; $r = r + 1$}
        \ENDWHILE%
    \end{algorithmic}
\end{algorithm}

%% file: ZMortonRecursion.tex
\begin{figure}
    \centering
    \resizebox{0.55\textwidth}{!}{%
        \begin{tikzpicture}
            % First Matrix (0,0) -> (4,4)
            \draw (0,0) rectangle (2,2);
            \draw (2,2) rectangle (4,4);
            \draw (0,2) rectangle (2,4);
            \draw (2,0) rectangle (4,2);
            \draw [dashed] (1,0) -- (1,4);
            \draw [dashed] (3,0) -- (3,4);
            \draw [dashed] (0,1) -- (4,1);
            \draw [dashed] (0,3) -- (4,3);
    
            % Second matrix (4.5,0) -> (8.5,4)
            \foreach \x in {0,...,3} {
                \foreach \y in {0,...,3} {
                    \draw (4.5+\x, \y) rectangle (5.5+\x, 1+\y);
                    \draw [dashed] (4.5+\x, 0.5+\y) -- (5.5+\x, 0.5+\y);
                    \draw [dashed] (5+\x, \y) -- (5+\x, 1+\y);
                }
            }
        
            % Hilbert curve in the first matrix
            \draw [line width=0.2mm, blue] (0.5,3.5) -- (1.5,3.5) -- (0.5,2.5) -- (1.5,2.5) -- (2.5,3.5) -- (3.5,3.5) -- (2.5,2.5) -- (3.5,2.5) -- (0.5,1.5) -- (1.5,1.5) -- (0.5,0.5) -- (1.5,0.5) -- (2.5,1.5) -- (3.5,1.5) -- (2.5,0.5) -- (3.5,0.5);
            \draw [line width=0.8mm, red] (1,3) -- (3,3) -- (1,1) -- (3,1);

            % Hilbert curve in the second matrix
            \draw [line width=0.2mm, blue, shift={(4.5, 0)}] (0.25, 3.75) -- (0.75, 3.75) -- (0.25, 3.25) -- (0.75, 3.25) -- (1.25, 3.75) -- (1.75, 3.75) -- (1.25, 3.25) -- (1.75, 3.25) -- (0.25, 2.75) -- (0.75, 2.75) -- (0.25, 2.25) -- (0.75, 2.25) -- (1.25, 2.75) -- (1.75, 2.75) -- (1.25, 2.25) -- (1.75, 2.25) -- (2.25, 3.75) -- (2.75, 3.75) -- (2.25, 3.25) -- (2.75, 3.25) -- (3.25, 3.75) -- (3.75, 3.75) -- (3.25, 3.25) -- (3.75, 3.25) -- (2.25, 2.75)-- (2.75, 2.75) -- (2.25, 2.25) -- (2.75, 2.25) -- (3.25, 2.75) -- (3.75, 2.75) -- (3.25, 2.25) -- (3.75, 2.25);
            \draw [line width=0.2mm, blue, shift={(4.5, -2)}] (0.25, 3.75) -- (0.75, 3.75) -- (0.25, 3.25) -- (0.75, 3.25) -- (1.25, 3.75) -- (1.75, 3.75) -- (1.25, 3.25) -- (1.75, 3.25) -- (0.25, 2.75) -- (0.75, 2.75) -- (0.25, 2.25) -- (0.75, 2.25) -- (1.25, 2.75) -- (1.75, 2.75) -- (1.25, 2.25) -- (1.75, 2.25) -- (2.25, 3.75) -- (2.75, 3.75) -- (2.25, 3.25) -- (2.75, 3.25) -- (3.25, 3.75) -- (3.75, 3.75) -- (3.25, 3.25) -- (3.75, 3.25) -- (2.25, 2.75)-- (2.75, 2.75) -- (2.25, 2.25) -- (2.75, 2.25) -- (3.25, 2.75) -- (3.75, 2.75) -- (3.25, 2.25) -- (3.75, 2.25);
            \draw [line width=0.2mm, blue, shift={(4.5, 0)}] (3.75, 2.25) -- (0.25, 1.75);
            \draw [line width=0.8mm, red, shift={(4.5, 0)}] (0.5,3.5) -- (1.5,3.5) -- (0.5,2.5) -- (1.5,2.5) -- (2.5,3.5) -- (3.5,3.5) -- (2.5,2.5) -- (3.5,2.5) -- (0.5,1.5) -- (1.5,1.5) -- (0.5,0.5) -- (1.5,0.5) -- (2.5,1.5) -- (3.5,1.5) -- (2.5,0.5) -- (3.5,0.5);
                    
        \end{tikzpicture}
    }
    \caption{Z-Morton curve projected on a two-by-two, four-by-four and eight-by-eight matrix.}\label{fig:ZMorton}
\end{figure}
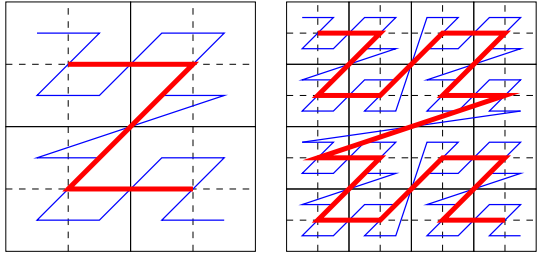

%% file: HilbertRecursion.tex
\begin{figure}
    \centering
    \resizebox{0.55\textwidth}{!}{%
        \begin{tikzpicture}
            % First Matrix (0,0) -> (4,4)
            \draw (0,0) rectangle (2,2);
            \draw (2,2) rectangle (4,4);
            \draw (0,2) rectangle (2,4);
            \draw (2,0) rectangle (4,2);
            \draw [dashed] (1,0) -- (1,4);
            \draw [dashed] (3,0) -- (3,4);
            \draw [dashed] (0,1) -- (4,1);
            \draw [dashed] (0,3) -- (4,3);
    
            % Second matrix (4.5,0) -> (8.5,4)
            \foreach \x in {0,...,3} {
                \foreach \y in {0,...,3} {
                    \draw (4.5+\x, \y) rectangle (5.5+\x, 1+\y);
                    \draw [dashed] (4.5+\x, 0.5+\y) -- (5.5+\x, 0.5+\y);
                    \draw [dashed] (5+\x, \y) -- (5+\x, 1+\y);
                }
            }
        
            % Hilbert curve in the first matrix
            \draw [line width=0.2mm, blue] (0.5,3.5) -- (1.5,3.5) -- (1.5,2.5) -- (0.5,2.5) -- (0.5,0.5) -- (1.5,0.5) -- (1.5,1.5) -- (2.5,1.5) -- (2.5,0.5) -- (3.5,0.5) -- (3.5,2.5) -- (2.5,2.5) -- (2.5,3.5) -- (3.5,3.5);
            \draw [line width=0.8mm, red] (1,3) -- (1,1) -- (3,1) -- (3,3);

            % Hilbert curve in the second matrix
            \draw [line width=0.2mm, blue, shift={(4.5, 0)}] (0.25, 3.75) -- (0.25, 3.25) -- (0.75, 3.25) -- (0.75, 3.75) -- (1.75, 3.75) -- (1.75, 3.25) -- (1.25, 3.25) -- (1.25, 2.75) -- (1.75, 2.75) -- (1.75, 2.25) -- (0.75, 2.25) -- (0.75, 2.75) -- (0.25, 2.75) -- (0.25, 1.75) -- (0.75, 1.75) -- (0.75, 1.25) -- (0.25, 1.25) -- (0.25, 0.25) -- (0.75, 0.25) -- (0.75, 0.75) -- (1.25, 0.75) -- (1.25, 0.25) -- (1.75, 0.25) -- (1.75, 1.25) -- (1.25, 1.25) -- (1.25, 1.75) -- (2.75, 1.75) -- (2.75, 1.25) -- (2.25, 1.25) -- (2.25, 0.25) -- (2.75, 0.25) -- (2.75, 0.75) -- (3.25, 0.75) -- (3.25, 0.25) -- (3.75, 0.25) -- (3.75, 1.25) -- (3.25, 1.25) -- (3.25, 1.75) -- (3.75, 1.75) -- (3.75, 2.75) -- (3.25, 2.75) -- (3.25, 2.25) -- (2.25, 2.25) -- (2.25, 2.75) -- (2.75, 2.75) -- (2.75, 3.25) -- (2.25, 3.25) -- (2.25, 3.75) -- (3.25, 3.75) -- (3.25, 3.25) -- (3.75, 3.25) -- (3.75, 3.75);
            \draw [line width=0.8mm, red] (4.5+0.5,3.5) -- (4.5+1.5,3.5) -- (4.5+1.5,2.5) -- (4.5+0.5,2.5) -- (4.5+0.5,0.5) -- (4.5+1.5,0.5) -- (4.5+1.5,1.5) -- (4.5+2.5,1.5) -- (4.5+2.5,0.5) -- (4.5+3.5,0.5) -- (4.5+3.5,2.5) -- (4.5+2.5,2.5) -- (4.5+2.5,3.5) -- (4.5+3.5,3.5);
                    
        \end{tikzpicture}
    }
    \caption{Hilbert curve projected on a two-by-two, four-by-four and eight-by-eight matrix.}\label{fig:Hilbert}
\end{figure}
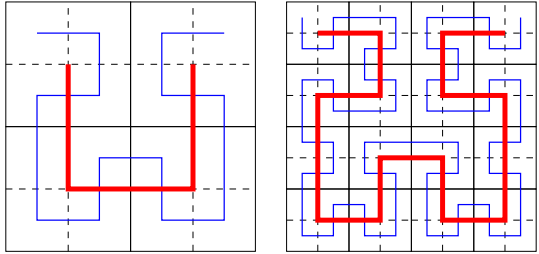

%% file: merge.tex
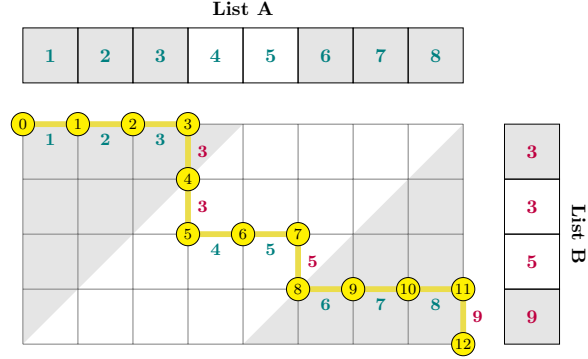
\begin{figure}
    \centering
    \resizebox{0.6\linewidth}{!}{%
        \begin{tikzpicture}
            % Matrix
            \draw[step=1, gray, very thin] (0,0) grid (8,4);

            % List A
            \node[align=center] at (4, 6.1) {\textbf{List A}};
            \draw[fill=gray, fill opacity=0.2] (0, 4.75) rectangle (1, 5.75) node[pos=0.5, teal, fill opacity=1] {\textbf{1}};
            \draw[fill=gray, fill opacity=0.2] (1, 4.75) rectangle (2, 5.75) node[pos=0.5, teal, fill opacity=1] {\textbf{2}};
            \draw[fill=gray, fill opacity=0.2] (2, 4.75) rectangle (3, 5.75) node[pos=0.5, teal, fill opacity=1] {\textbf{3}};
            \draw                              (3, 4.75) rectangle (4, 5.75) node[pos=0.5, teal]                 {\textbf{4}};
            \draw                              (4, 4.75) rectangle (5, 5.75) node[pos=0.5, teal]                 {\textbf{5}};
            \draw[fill=gray, fill opacity=0.2] (5, 4.75) rectangle (6, 5.75) node[pos=0.5, teal, fill opacity=1] {\textbf{6}};
            \draw[fill=gray, fill opacity=0.2] (6, 4.75) rectangle (7, 5.75) node[pos=0.5, teal, fill opacity=1] {\textbf{7}};
            \draw[fill=gray, fill opacity=0.2] (7, 4.75) rectangle (8, 5.75) node[pos=0.5, teal, fill opacity=1] {\textbf{8}};

            % List B
            \node[align=center, rotate=-90] at (10.1, 2) {\textbf{List B}};
            \draw[fill=gray, fill opacity=0.2] (8.75, 0) rectangle (9.75, 1) node[pos=0.5, purple, fill opacity=1] {\textbf{9}};
            \draw                              (8.75, 1) rectangle (9.75, 2) node[pos=0.5, purple]                 {\textbf{5}};
            \draw                              (8.75, 2) rectangle (9.75, 3) node[pos=0.5, purple]                 {\textbf{3}};
            \draw[fill=gray, fill opacity=0.2] (8.75, 3) rectangle (9.75, 4) node[pos=0.5, purple, fill opacity=1] {\textbf{3}};

            % Diagonals + coloring
            \draw[draw opacity=0, fill=gray, fill opacity=0.2] (0, 0) -- (4, 4) -- (0, 4);
            \draw[draw opacity=0, fill=gray, fill opacity=0.2] (4, 0) -- (8, 4) -- (8, 0);

            % Path
            \draw[yellow!80!gray, line width=3pt] (0,4) -- node[below, teal, font=\small]   {\textbf{1}} (1,4);
            \draw[yellow!80!gray, line width=3pt] (1,4) -- node[below, teal, font=\small]   {\textbf{2}} (2,4);
            \draw[yellow!80!gray, line width=3pt] (2,4) -- node[below, teal, font=\small]   {\textbf{3}} (3,4);
            \draw[yellow!80!gray, line width=3pt] (3,4) -- node[right, purple, font=\small] {\textbf{3}} (3,3);
            \draw[yellow!80!gray, line width=3pt] (3,3) -- node[right, purple, font=\small] {\textbf{3}} (3,2);
            \draw[yellow!80!gray, line width=3pt] (3,2) -- node[below, teal, font=\small]   {\textbf{4}} (4,2);
            \draw[yellow!80!gray, line width=3pt] (4,2) -- node[below, teal, font=\small]   {\textbf{5}} (5,2);
            \draw[yellow!80!gray, line width=3pt] (5,2) -- node[right, purple, font=\small] {\textbf{5}} (5,1);
            \draw[yellow!80!gray, line width=3pt] (5,1) -- node[below, teal, font=\small]   {\textbf{6}} (6,1);
            \draw[yellow!80!gray, line width=3pt] (6,1) -- node[below, teal, font=\small]   {\textbf{7}} (7,1);
            \draw[yellow!80!gray, line width=3pt] (7,1) -- node[below, teal, font=\small]   {\textbf{8}} (8,1);
            \draw[yellow!80!gray, line width=3pt] (8,1) -- node[right, purple, font=\small] {\textbf{9}} (8,0);

            % Coordinate nodes
            \draw[fill=yellow] (0,4) circle (6pt) node[font=\footnotesize] {0};
            \draw[fill=yellow] (1,4) circle (6pt) node[font=\footnotesize] {1};
            \draw[fill=yellow] (2,4) circle (6pt) node[font=\footnotesize] {2};
            \draw[fill=yellow] (3,4) circle (6pt) node[font=\footnotesize] {3};
            \draw[fill=yellow] (3,3) circle (6pt) node[font=\footnotesize] {4};
            \draw[fill=yellow] (3,2) circle (6pt) node[font=\footnotesize] {5};
            \draw[fill=yellow] (4,2) circle (6pt) node[font=\footnotesize] {6};
            \draw[fill=yellow] (5,2) circle (6pt) node[font=\footnotesize] {7};
            \draw[fill=yellow] (5,1) circle (6pt) node[font=\footnotesize] {8};
            \draw[fill=yellow] (6,1) circle (6pt) node[font=\footnotesize] {9};
            \draw[fill=yellow] (7,1) circle (6pt) node[font=\footnotesize] {10};
            \draw[fill=yellow] (8,1) circle (6pt) node[font=\footnotesize] {11};
            \draw[fill=yellow] (8,0) circle (6pt) node[font=\footnotesize] {12};
        \end{tikzpicture}
    }
    \caption{Visual representation of parallel merge-sort using three threads\protect\footnotemark.}\label{fig:SPMV-MergePath}
\end{figure}

%% file: TestMatrices.tex
\begin{table}
    \centering
    \small
    \caption{Test matrices with their relevant properties\protect\footnotemark. All test matrices are square. Max and variance denote the maximum number of nonzero elements per row and the variance of nonzero elements per row respectively. The matrices are split in a class with low density (top), and higher density (bottom).}\label{tab:Matrices}
    \begin{tabular}{l S[table-format=9.0] S[table-format=9.0] S[table-format=1.2e2] S[table-format=9.0] S[table-format=1.2e2]}
        Matrix & \head{size} & \head{nonzeros} & \head{density} & \head{max} & \head{variance}\\ 
        \midrule
        \texttt{kmer\_P1a}      & 139 353 211 & 297 829 984  & 1.53E-8 & 40          & 4.06E-1\\
        \texttt{mawi\_0130}     & 128 568 730 & 270 234 840  & 1.63E-8 & 119 191 841 & 1.11E+8\\
        \texttt{europe\_osm}    & 50 912 018  & 108 109 320  & 4.17E-8 & 13          & 2.30E-1\\
        \texttt{LHH}            & 50 000 000  & 171 108 095  & 6.84E-8 & 144         & 6.25E+1\\
        \texttt{road\_usa}      & 23 947 347  & 57 708 624   & 1.01E-7 & 9           & 8.63E-1\\
        \texttt{hugebubbles}    & 21 198 119  & 63 580 358   & 1.41E-7 & 3           & 6.60E-4\\
        \texttt{hugetrace}      & 16 002 413  & 47 997 626   & 1.87E-7 & 3           & 6.00E-4\\
        \texttt{wb\_edu}        & 9 845 725   & 57 156 537   & 5.89E-7 & 3 841       & 4.12E+2\\        
        \texttt{uk-2002}        & 18 520 486  & 298 113 762  & 8.69E-7 & 2 450       & 7.58E+2\\
        \midrule
        \texttt{ljournal-2008}  & 5 363 260   & 79 023 142   & 2.77E-6 & 2 469       & 1.37E+3\\
        \texttt{LiveJournal1}   & 4 847 571   & 68 993 773   & 2.94E-6 & 20 293      & 1.30E+3\\
        \texttt{indochina-2004} & 7 414 866   & 194 109 311  & 3.53E-6 & 6 985       & 4.66E+4\\
        \texttt{cage15}         & 5 154 859   & 99 199 551   & 3.73E-6 & 47          & 3.29E+1\\
        \texttt{HHH}            & 3 120 000   & 38 286 436   & 3.93E-6 & 537         & 1.14E+3\\
        \texttt{com-Orkut}      & 3 072 441   & 234 370 166  & 2.48E-5 & 33 313      & 2.40E+4\\
        \texttt{kron}           & 2 097 152   & 182 082 942  & 4.14E-5 & 213 905     & 5.71E+5\\
    \end{tabular}
\end{table}

%% file: TestMachines.tex
\begin{table}
    \centering
    \small
    \caption{Test machines with their relevant properties. Each memory stick is placed such that it is the only stick in the memory channel. If no thread count is reported, hyperthreading is turned off.}\label{tab:Machines}
    \begin{tabular}{r c c c}
        Name & CPU & RAM\\
        \midrule
        Sapphire Rapids & $2 \times 48$ cores @ \SI{2.1}{\giga\hertz} & $16 \times 16$ \SI{}{\giga\byte}, \SI{4800}{\mega\hertz}, DDR5\\
        Ice Lake NUMA     & $2 \times 36$ cores @ \SI{2.4}{\giga\hertz} & $16 \times 16$ \SI{}{\giga\byte}, \SI{3200}{\mega\hertz}, DDR4\\
        Ice Lake UMA     & $36$ cores @ \SI{2.4}{\giga\hertz} & $16 \times 16$ \SI{}{\giga\byte}, \SI{3200}{\mega\hertz}, DDR4\\
        Cascade Lake  & $18$ cores, $36$ threads @ \SI{3}{\giga\hertz} & $2 \times 16$ \SI{}{\giga\byte}, \SI{3200}{\mega\hertz}, DDR4\\
    \end{tabular}
\end{table}

%% file: AggregateL.tex
\begin{table}
    \centering
    \small
    \caption{Average best parallel speedup for test matrices with low density on different machines. The algorithm which obtains the highest speedup for each machine is indicated in bold. The number of threads for which the result is obtained is in parentheses. The names of the newly developed methods are in bold.}\label{tab:AggregateResultsL}
    \begin{tabular}{r r@{ }l r@{ }l r@{ }lr@{ }l }
        Methods & \multicolumn{2}{c}{SR} & \multicolumn{2}{c}{IL NUMA} & \multicolumn{2}{c}{IL UMA} & \multicolumn{2}{c}{CL}\\
        \midrule
        ParCRS           & 42.2          & (96)          & 26.4          & (72)          & 18.8          & (36)          & 2.9          & (10) \\
        Merge            & 43.6          & (96)          & \textbf{33.1} & \textbf{(72)} & 18.0          & (36)          & 2.8          & (12) \\
        CSB              & 29.4          & (96)          & 22.7          & (72)          & 18.9          & (36)          & 3.3          & (12) \\
        \textbf{CSBH}    & 30.4          & (96)          & 23.5          & (72)          & \textbf{19.1} & \textbf{(36)} & \textbf{3.4} & \textbf{(12)} \\
        BCOH             & 45.8          & (96)          & 27.7          & (72)          & 13.7          & (36)          & 2.1          & (6) \\
        \textbf{BCOHC}   & 49.6          & (96)          & 29.1          & (72)          & 14.5          & (36)          & 2.1          & (6) \\
        \textbf{BCOHCH}  & \textbf{49.7} & \textbf{(96)} & 28.8          & (72)          & 14.2          & (36)          & 2.0          & (6) \\
        \textbf{BCHOCHP} & 26.7          & (96)          & 17.8          & (72)          & 11.2          & (36)          & 2.0          & (8) \\
        \textbf{MergeB}  & 22.6          & (56)          & 19.6          & (72)          & 15.0          & (36)          & 3.1          & (8) \\
        \textbf{MergeBH} & 23.3          & (56)          & 21.0          & (72)          & 15.6          & (36)          & 3.1          & (8) \\
    \end{tabular}
\end{table}

%% file: AggregateH.tex
\begin{table}
    \centering
    \small
    \caption{As Table~\ref{tab:AggregateResultsL}, for test matrices with higher density.}\label{tab:AggregateResultsH}
    \begin{tabular}{r r@{ }l r@{ }l r@{ }l r@{ }l}
        Methods & \multicolumn{2}{c}{SR} & \multicolumn{2}{c}{IL NUMA} & \multicolumn{2}{c}{IL UMA} & \multicolumn{2}{c}{CL}\\
        \midrule
        ParCRS           & 55.2          & (96)          & 38.7          & (72)          & \textbf{25.8} & \textbf{(36)} & \textbf{4.5} & \textbf{(18)}\\
        Merge            & 71.3          & (96)          & 48.3          & (72)          & 24.4          & (36)          & \textbf{4.5} & \textbf{(18)}\\
        CSB              & 33.7          & (72)          & 24.4          & (68)          & 20.5          & (36)          & 4.3          & (12)\\
        \textbf{CSBH}    & 37.1          & (96)          & 26.5          & (72)          & 21.3          & (36)          & 4.3          & (12)\\
        BCOH             & 59.5          & (96)          & 35.7          & (72)          & 18.0          & (36)          & 4.4          & (18)\\
        \textbf{BCOHC}   & 81.9          & (96)          & 49.6          & (72)          & 24.4          & (36)          & 4.1          & (18)\\
        \textbf{BCOHCH}  & \textbf{84.6} & \textbf{(96)} & \textbf{53.8} & \textbf{(72)} & 25.6          & (36)          & 4.1          & (18)\\
        \textbf{BCHOCHP} & 72.1          & (96)          & 45.3          & (72)          & 23.6          & (36)          & 4.0          & (18)\\
        \textbf{MergeB}  & 33.3          & (64)          & 22.0          & (64)          & 14.8          & (36)          & 4.4          & (18)\\
        \textbf{MergeBH} & 37.1          & (64)          & 26.7          & (64)          & 17.3          & (36)          & \textbf{4.5} & \textbf{(16)}\\
        \end{tabular}
\end{table}

%% file: MawiSpeedup.tex
\begin{table}
    \centering
    \small
    \caption{Best parallel speedup for \texttt{mawi\_0130}. The algorithm which obtains the highest speedup for each machine is indicated in bold. The number of threads for which the result is obtained is in parentheses. The names of the newly developed methods are in bold.}\label{tab:MawiSpeedup}
    \begin{tabular}{r r@{ }l r@{ }l r@{ }lr@{ }l }
        Methods & \multicolumn{2}{c}{SR} & \multicolumn{2}{c}{IL NUMA} & \multicolumn{2}{c}{IL UMA} & \multicolumn{2}{c}{CL}\\
        \midrule
        ParCRS           &  2.7          & (56)          &  3.8          & (72)          &  2.3          & (16)          & 1.7          & (6) \\
        Merge            & \textbf{23.5} & \textbf{(88)} & \textbf{24.2} & \textbf{(64)} & 10.7          & (36)          & 2.1          & (14) \\
        CSB              & 16.8          & (96)          & 18.8          & (72)          & \textbf{11.9} & \textbf{(36)} & \textbf{2.3} & \textbf{(6)} \\
        \textbf{CSBH}    & 16.7          & (96)          & 18.8          & (68)          & 11.8          & (36)          & \textbf{2.3} & \textbf{(6)} \\
        BCOH             &  1.9          & (52)          &  1.5          & (40)          &  1.8          & (16)          & 1.7          & (6) \\
        \textbf{BCOHC}   &  1.9          & (52)          &  1.6          & (40)          &  1.8          & (16)          & 1.6          & (6) \\
        \textbf{BCOHCH}  &  1.9          & (52)          &  1.6          & (40)          &  1.8          & (16)          & 1.6          & (6) \\
        \textbf{BCHOCHP} &  1.6          & (52)          &  1.3          & (40)          &  1.5          & (16)          & 1.5          & (8) \\
        \textbf{MergeB}  &  3.1          & (64)          &  3.0          & (56)          &  3.3          & (24)          & 2.2          & (16) \\
        \textbf{MergeBH} &  3.0          & (64)          &  3.0          & (56)          &  3.3          & (24)          & 2.1          & (16) \\
    \end{tabular}
\end{table}

%% file: AggregateSpeedup.tex
% Figures are obtained using figsize (5.25,3.75) and legend ncol=2 lower right

\begin{figure}
    \centering
    \begin{subfigure}{0.45\textwidth}
        \includegraphics[width=\textwidth]{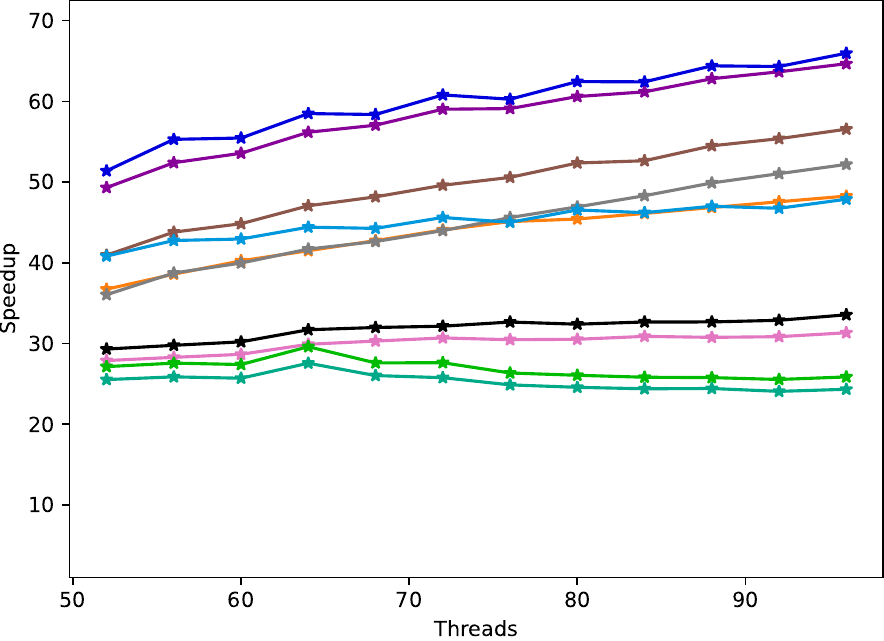}
        \caption{Sapphire Rapids}
    \end{subfigure}
    \hspace{0.05\textwidth}
    \begin{subfigure}{0.45\textwidth}
        \includegraphics[width=\textwidth]{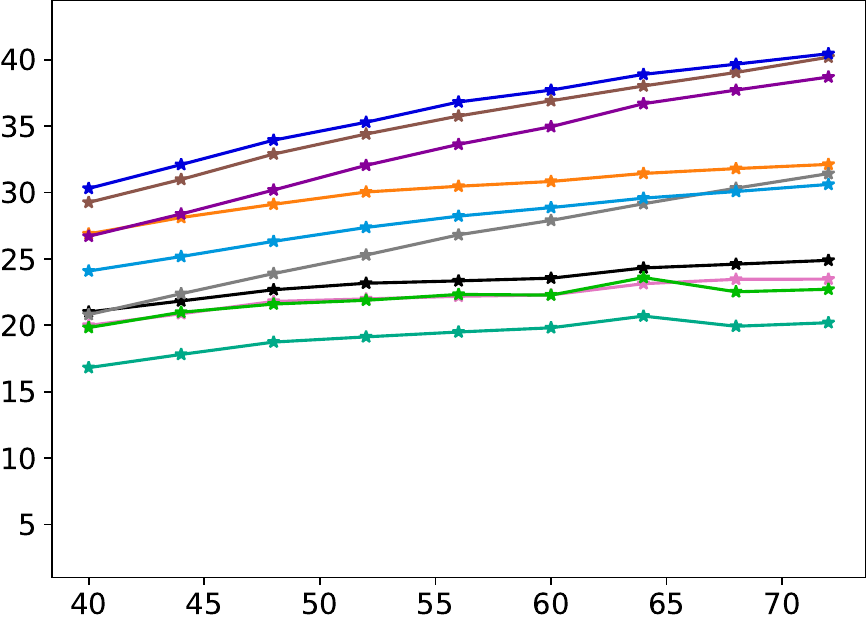}
        \caption{Ice Lake NUMA}
    \end{subfigure}
    \begin{subfigure}{0.45\textwidth}
        \includegraphics[width=\textwidth]{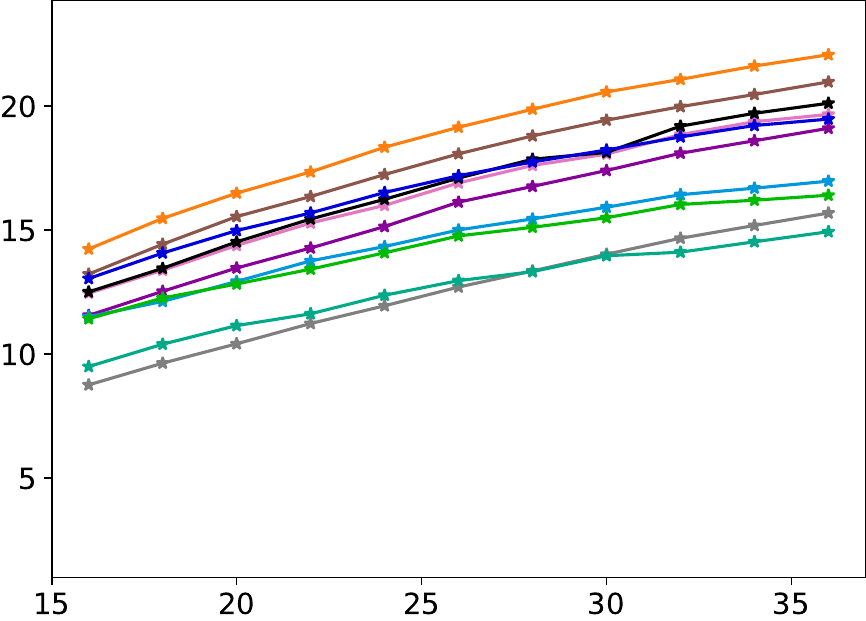}
        \caption{Ice Lake UMA}
    \end{subfigure}
    \hspace{0.05\textwidth}
    \begin{subfigure}{0.45\textwidth}
        \includegraphics[width=\textwidth]{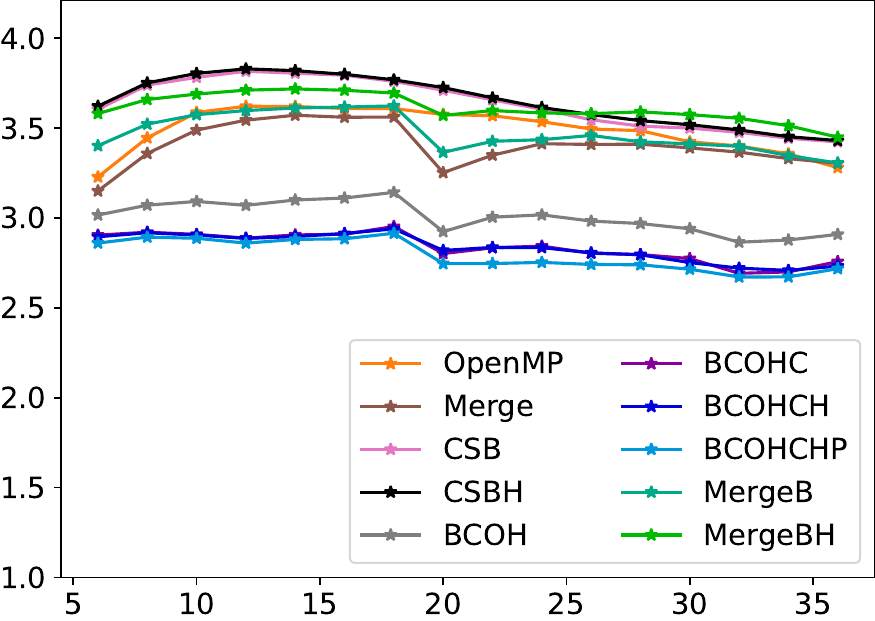}
        \caption{Cascade Lake}\label{fig:AvgParSpeedupEclipse}
    \end{subfigure}
    \caption{Average parallel speedup for different machines. The horizontal axis depicts the number of threads used. The vertical axis depicts the speedup.}\label{fig:AvgParSpeedup}
\end{figure}

%% file: ConversionL.tex
\begin{table}
    \centering
    \small
    \caption{Average conversion time from the triplet storage format to the storage format needed by a specific method for the low density test matrices. The conversion time is displayed as the amount of ParCRS SpMV multiplications. The conversion is performed using the amount of threads that provided the most optimal parallel speedup for the SpMV multiplication. The names of the newly developed methods are in bold}\label{tab:ConversionL}
    \begin{tabular}{r S[table-format=3.1] S[table-format=3.1] S[table-format=3.1] S[table-format=3.1]}
        Methods & \head{SR} & \head{IL NUMA} & \head{IL UMA} & \head{CL}\\
        \midrule
        ParCRS           & 79.1  & 28.4  & 24.0  &  7.8  \\
        Merge            & 74.8  & 27.5  & 24.0  &  7.6  \\
        CSB              & 67.1  & 29.1  & 33.6  & 10.9  \\
        \textbf{CSBH}    & 260.9 & 213.5 & 289.4 & 69.6  \\
        BCOH             & 184.1 & 149.3 & 189.4 & 74.4  \\
        \textbf{BCOHC}   & 175.7 & 142.3 & 178.1 & 69.0  \\
        \textbf{BCOHCH}  & 412.4 & 376.9 & 507.0 & 204.1 \\
        \textbf{BCHOCHP} & 413.0 & 377.2 & 506.1 & 155.3 \\
        \textbf{MergeB}  & 63.5  & 19.1  & 21.0  & 10.1  \\
        \textbf{MergeBH} & 352.5 & 211.8 & 289.8 & 103.2 \\
    \end{tabular}
\end{table}

%% file: ConversionH.tex
\begin{table}
    \centering
    \small
    \caption{As Table~\ref{tab:ConversionL}, for test matrices with higher density.}\label{tab:ConversionH}
    \begin{tabular}{r S[table-format=3.1] S[table-format=3.1] S[table-format=3.1] S[table-format=3.1]}
        Methods & \head{SR} & \head{IL NUMA} & \head{IL UMA} & \head{CL}\\
        \midrule
        ParCRS           & 132.5 & 46.2  & 38.3  & 14.7  \\
        Merge            & 130.7 & 44.6  & 38.5  & 14.7  \\
        CSB              & 132.1 & 70.6  & 75.9  & 23.6  \\
        \textbf{CSBH}    & 490.1 & 527.7 & 683.0 & 163.4 \\
        BCOH             & 291.6 & 293.9 & 353.5 & 54.0  \\
        \textbf{BCOHC}   & 287.3 & 292.4 & 352.4 & 54.1  \\
        \textbf{BCOHCH}  & 646.5 & 736.6 & 945.1 & 143.3 \\
        \textbf{BCHOCHP} & 646.2 & 737.7 & 941.8 & 143.3 \\
        \textbf{MergeB}  & 117.8 & 47.0  & 44.8  & 14.6  \\
        \textbf{MergeBH} & 626.4 & 589.1 & 683.5 & 124.5 \\
    \end{tabular}
\end{table}